\documentclass[sigconf]{acmart}

\setcopyright{none}
\AtBeginDocument{%
  }

\acmYear{2025}
\acmDOI{}
\acmConference[WebSci PhD Symposium '25]{WebSci PhD Symposium '25}{May 20--23, 2025}{New Brunswick, NJ}
\acmISBN{}

\begin{document}

\title{Persuasion and Safety in the Era of Generative AI}

\author{Haein Kong}
\affiliation{%
  \institution{Rutgers University}
  \city{New Brunswick}
  \state{NJ}
  \country{USA}
}
\email{haein.kong@rutgers.edu}

\renewcommand{\shortauthors}{Kong}

% \begin{abstract}

% \end{abstract}







\begin{abstract}
As large language models (LLMs) achieve advanced persuasive capabilities, concerns about their potential risks have grown. The EU AI Act prohibits AI systems that use manipulative or deceptive techniques to undermine informed decision-making, highlighting the need to distinguish between rational persuasion, which engages reason, and manipulation, which exploits cognitive biases. My dissertation addresses the lack of empirical studies in this area by developing a taxonomy of persuasive techniques, creating a human-annotated dataset, and evaluating LLMs’ ability to distinguish between these methods. This work contributes to AI safety by providing resources to mitigate the risks of persuasive AI and fostering discussions on ethical persuasion in the age of generative AI.
\end{abstract}

%Generative AI has advanced rapidly in recent years. 
% Persuasive AI contains risks in that it can be misused to manipulate humans or spread misinformation. 
% to perform many tasks that require intelligence. It has been found that LLMs have
% As more intelligent and persuasive AI systems emerge, the risks in those systems have also been recognized. 
\maketitle
\section{Introduction}
Large language models (LLMs) have shown their capabilities in various tasks that require an understanding of the human mind and behavior, such as the theory of mind \cite{kosinski2023theory}, emotion detection \cite{kong2024ru}, etc. Moreover, recent research has found that LLMs are capable of generating high-quality persuasive content for diverse domains \cite{bai2023artificial, rogiers2024PersuasionSurvey}, which can eventually influence humans by changing their thoughts or behaviors. Recent growth in the persuasive skills of AI models in successive models \cite{Anthropic2024persuasion} suggests a strong growth curve in the persuasion abilities of future AI models. 

Researchers have started to investigate the harms of persuasive AI as LLMs have obtained advanced persuasion skills \cite{bengio2025international, el2024mechanism}. One of the efforts is to distinguish the types of persuasion. In a standard taxonomy \cite{el2024mechanism, jones2024lies}, persuasion is divided into two different types, rational persuasion and manipulation, based on their persuasion mechanisms. Researchers have argued the harms of manipulation as its nature relies on cognitive shortcuts and heuristics --which often bypass one's conscious awareness \cite{el2024mechanism}. Given that manipulation can harm one's autonomy, persuasive AI that relies on manipulation can be considered unsafe. 

Hence, it becomes important to distinguish safe versus unsafe persuasion in this age of generative AI. One criterion to classify safe persuasion can be based on the taxonomy mentioned earlier: rational persuasion versus manipulation. However, previous NLP research on persuasion underexplored the two mechanisms. For instance, previous works often used persuasion on a broad level and did not pay much attention to the different mechanisms of persuasion techniques \cite{PersuasionForGood, DailyPersuasion}. This may be because AI persuasion is a relatively new concept, and there is a lack of awareness of the two subtypes of persuasion.

Therefore, my research aims to address this gap. Specifically, my dissertation will conduct studies that include 1) proposing a taxonomy differentiating persuasion techniques that belong to rational persuasion or manipulation, 2) building a dataset with human annotation for the above, and 3) evaluating the current LLMs' capability to differentiate these concepts. We also plan to build upon this work by suggesting novel ways that can improve the classification performance of LLMs on our dataset. 

The following sections start with related works in persuasive AI, including the current findings, risks, and theoretical backgrounds. Then, the research plan will be presented, and the plan for the studies in my dissertation will be described.

\section{Related Works}

% 2.1 
\subsection{Persuasive AI}
LLM-powered persuasion has been studied in various application domains, such as public health, politics, and e-commerce \cite{rogiers2024PersuasionSurvey}. Recent research findings have shown that LLMs are capable of generating content as persuasive as humans \cite{Anthropic2024persuasion, rogiers2024PersuasionSurvey, bai2023artificial}. For example, \citet{bai2023artificial} found that LLM-generated messages on policy issues are as persuasive as human-generated messages. Similarly, \citet{Anthropic2024persuasion} showed that the persuasiveness score of the arguments generated by Claude 3 Opus is not significantly different from the persuasiveness score of human-generated arguments. Many research findings suggest that LLMs' persuasive capabilities are on par with humans or even exceed human performance \cite{rogiers2024PersuasionSurvey}. There is a study that found a scaling trend in persuasion skills, showing that successive model generations tend to have higher persuasion scores than the previous models \cite{Anthropic2024persuasion}. This suggests that LLMs can potentially match or exceed human persuasion skills soon.  

% 2.2
\subsection{Risk in Persuasive AI}
As LLMs achieve considerable persuasion skills, the risk in persuasive AI has also been recognized \cite{hendrycks2023overview, bengio2025international}. Researchers have pointed out that persuasive AI could be used to generate personalized disinformation \cite{hendrycks2023overview}, manipulate public opinion \cite{bengio2025international}, and be misused by malicious actors such as criminals, governments, and others \cite{jones2024lies}. Tech industries are also aware of the risks of persuasive AI. For example, OpenAI defined persuasion as one of the risks and evaluated GPT 4o as a medium risk for political persuasion through text \cite{OpenAI2024o1}. Also, Anthropic, a company that developed Claude, tries to prevent their systems from being used to generate persuasive content for harmful purposes \cite{Anthropic2024persuasion}. 

In addition, the EU AI Act contains regulations about prohibiting certain AI systems, including those manipulating human decisions or exploiting vulnerabilities \cite{EuAiAct}. The paragraph 1 of Article 5 mentions that ``(a) the placing on the market, the putting into service or the use of an AI system that deploys subliminal techniques beyond a person’s consciousness or purposefully manipulative or deceptive techniques, with the objective, or the effect of materially distorting the behaviour of a person or a group of persons by appreciably impairing their ability to make an informed decision, thereby causing them to take a decision that they would not have otherwise taken in a manner that causes or is reasonably likely to cause that person, another person or group of persons significant harm'' \cite{EuAiAct}. While there are arguments about the interpretations and definitions of subliminal, manipulative, and/or deceptive techniques \cite{zhong2024regulating}, this regulation suggests that the AI systems that can harm one's cognitive autonomy and informed decision contain high risk. 

% 2.3
\subsection{Type of Persuasion}
Attempts to differentiate the type of persuasion as the risk in persuasive AI have gained academic attention. As mentioned in the previous section, recent works investigate the different types of persuasion based on the persuasion mechanism \cite{jones2024lies, el2024mechanism}. First, persuasion is one of influence, aiming to make an influence on others' decision-making \cite{jones2024lies, el2024mechanism}. Then, there are two subtypes of persuasion, which are rational persuasion and manipulation. Rational persuasion refers to ``influencing a person’s thoughts, attitudes, or behaviours through reason, evidence, and sound argument, along with intent, on the part of the message sender, to achieve these goals through their communication,'' while manipulation refers to ``intentionally and covertly influencing [someone’s] decision-making, by targeting and exploiting their decision-making vulnerabilities'' \cite{el2024mechanism, susser2019technology}. They argue that manipulation contains process harms in their usage of cognitive shortcuts, which can diminish humans' cognitive autonomy \cite{el2024mechanism}. 

\subsection{Theoretical Background}\label{sec:theory}
Understanding how humans process information is closely related to the aforementioned mechanism of persuasion. To better understand the connection between persuasion and human cognition, this section will briefly introduce the two theories that explain how humans access information and/or persuasion: Dual process theory and the Heuristics and Systematic Model. 

The dual process theory posits that there are two distinct sets of cognitive processes underlying human thinking and reasoning, which are referred to as System 1 and System 2 \cite{kahneman2011thinking}. System 1 is an implicit, fast, heuristic, automatic, nonverbal, and associative system, while System 2 is a conscious, slow, deliberate, rational, and explicit system \cite{pinder2018digital}. While System 1 often operates unconsciously and effortlessly, System 2 requires conscious awareness and attention as it engages in solving complex tasks \cite{zhong2024regulating}.
% From the perspective of dual process theory, human behavior is an outcome of both System 1 and 2 \cite{pinder2018digital}. Dual process theory is considered as a family of theories rather than one definitive theory as there are many theories with dual process approaches evolved in different research fields \cite{pinder2018digital}. 

Similarly, the Heuristics and Systematic Model (HSM), widely used in persuasion research, has dual-processing assumptions  \cite{bohner1995interplay, bohner2008information, gass2022persuasion}. This model assumes that there are two distinct modes when processing persuasive information: heuristic processing and systematic processing. It suggests that heuristic processing requires little cognitive resource and less effort, but systematic processing requires comprehensive processing of information and cognitive efforts \cite{bohner1995interplay}. HSM also proposes the least effort principle, that people prefer less effort and only spend cognitive resources when their interests are engaged \cite{bohner1995interplay}. It suggests that the default processing strategy is the heuristic mode, as it requires lower cognitive efforts than systematic processing. 

Given the two distinct mechanisms of information processing in human cognition, the aforementioned types of persuasion can be assumed to follow different systems or modes. For example, manipulation depends on the System 1/heuristics mode, while rational persuasion relies on the System 2/systematic mode. Based on this assumption, we argue that the persuasion that targets the fast and heuristic process of human cognition is risky in the era of generative AI. 

\section{Research Plan}
Based on the literature review, we argue that differentiating between the two types of persuasion, rational persuasion versus manipulation, is an important task in this era of generative AI. To classify whether the persuasion is rational or manipulative, the datasets with ground truths (human-annotated labels of rational persuasion and manipulation) are needed to build a classification model and measure the performance. However, we found there is no existing dataset built on this classification in the literature. Also, there is a lack of empirical studies that investigate whether the persuasion attempt is rational persuasion or manipulation.

Therefore, my dissertation aims to fill this gap by offering a taxonomy, creating a dataset based on the proposed taxonomy, and evaluating the performance of current state-of-the-art LLMs on this task. After completing the first part, we can build on it by designing new methods that improve the classification performance of LLMs to automatically detect unsafe persuasion attempts. 
% My dissertation can be divided into two parts, which are presented as Stage 1 and Stage 2 in the following section.

\subsection{Stage 1. Building a dataset and evaluating baseline models}

The first part is the construction of a multi-step taxonomy for the two different persuasion mechanisms: rational persuasion and manipulation. There is preliminary work on the taxonomy construction \cite{el2024mechanism}, but we aim to build a taxonomy that can finally be used to build a new dataset. Therefore, the main purpose of this taxonomy is to classify the examples of persuasive techniques for each persuasion category. This taxonomy will include persuasive techniques that reflect the differences between rational persuasion and manipulation. This will also include the techniques frequently used in the previous NLP studies. 

The next part is to build a new dataset based on the taxonomy. While there are several studies and datasets regarding persuasion in NLP literature \cite{PersuasionForGood, WinningArguments, DailyPersuasion}, there is no existing dataset that contains the labels of rational persuasion and manipulation. Thus, we will fill this gap by offering a human-annotated dataset. We plan to use a publicly available dataset as a source dataset. Specifically, we consider using the WinningArguments dataset \cite{WinningArguments}, a dataset of the subreddit \textit{r/ChangeMyView}. Since this dataset contains a large number of comments, we will perform data cleaning and preprocessing to filter out the relevant comments that contain the intention of persuasion. 

Then, we will manually annotate the filtered dataset based on the taxonomy. The research team will conduct this human annotation process due to the complex and nuanced nature of the topic. Reading comments and determining whether each comment is rational persuasion or manipulation requires cognitive resources and effort. To make our dataset credible and reliable, my coauthors and I will have tutorial sessions with a pilot test before starting the annotation. We aim to build a final dataset with roughly 1,000 comments for rational persuasion and manipulation, respectively.

Finally, we will evaluate the performance of the state-of-the-art LLMs (e.g., GPT, Llama, etc) on our dataset. The main task is to test whether the LLMs can classify the two different persuasion types (rational persuasion vs manipulation) and the persuasion techniques under each category. This evaluation will include basic methods such as zero-shot and few-shot prompting. 

\subsection{Stage 2. Improving LLMs' performance}

The next part is a follow-up study of stage 1. We aim to suggest a novel method to improve LLMs' classification performance. This can be done in multiple ways. For example, we can design new prompt engineering methods, fine-tuning, or frameworks that can improve classification performance. 
%After evaluating the baseline model performance, w

One of the potential ways is to develop a new prompting strategy by using cognitive theories. There are several theories that explain the mechanisms of human information processing (See Section \ref{sec:theory}). A new prompting can be developed based on these theories. For instance, we can add prior steps, such as asking whether the comment will be processed in System 1 or 2, and use that information to determine whether it is manipulation or rational persuasion. This approach will show how cognitive theories for human behavior can be used to improve the LLMs' capabilities. This part will be made more concrete after completing Stage 1. 
\section{Conclusion}
My dissertation will contribute to AI safety research, especially for the safety of persuasive AI. My work highlights the different types of persuasion and aims to offer a taxonomy, dataset, and evaluation of current LLMs. We will follow this by designing new approaches to automatically flag unsafe persuasion attempts. We expect that our work will facilitate research and discussions about the safety of persuasion and the importance of distinguishing rational persuasion and manipulation. Future research can use our dataset as a benchmark dataset, improve our taxonomy by adding more persuasion subtechniques, or suggest methods that can improve the detection performance of LLMs. Ultimately, these advancements will support the creation of safer use of AI in the field of persuasion. 

% We expect This task can be used to determine the safety of the persuasion driven by AI. This makes it hard to differentiate these two persuasion concepts, which is one of the essential tasks to protect humans from AI-driven manipulation. 

\begin{acks}
I appreciate Prof. Vivek Singh and Prof. Ruixiang Tang for their guidance and productive feedback on this research.
\end{acks}

\bibliographystyle{ACM-Reference-Format}
\bibliography{reference}
% \input{output.bbl}

% \appendix









\end{document}